\documentclass[9pt]{article}
\usepackage{amsfonts}
\usepackage{pstricks}
\usepackage{pst-node}
\usepackage{epsfig}
\usepackage{epsfig}
\usepackage[latin5]{inputenc}
\usepackage[T1]{fontenc}
\usepackage{lipsum}
\usepackage{amsmath}
\usepackage{authblk}

\begin{document}
                \def\ba{\begin{eqnarray}}
                \def\ea{\end{eqnarray}}
                \def\w{\wedge}
                \def\d{\mbox{d}}
                \def\D{\mbox{D}}

\begin{titlepage}
\title{ Variational Field Equations of a Majorana Neutrino Coupled to Einstein's Theory of General Relativity  }
\author{Tekin Dereli${}^{1,2,}$\footnote{tdereli@ku.edu.tr, tekindereli@maltepe.edu.tr} , Yorgo \c{S}eniko\u{g}lu${}^{1,}$\footnote{yorgosenikoglu@maltepe.edu.tr}}
\date{%
    ${}^{1}$ \small Department of Basic Sciences, Faculty of Engineering and Natural Sciences, \\Maltepe University, 34857 Maltepe,\.{I}stanbul, Turkey\\%
    ${}^{2}$ \small Department of Physics, Ko\c{c} University, 34450 Sar{\i}yer, Istanbul,Turkey\\[2ex]%
    \today
}
\maketitle



\begin{abstract}

\noindent  A consistent variational derivation of the Majorana 4-spinor field equations coupled to Einstein's theory of gravitation is given.
The equivalence of the first and the second order variational field equations is explicitly demonstrated. The Lagrange multiplier 2-forms we use turn out to be
precisely the Belinfante-Rosenfeld 2-forms that are needed to symmetrize the canonical energy-momentum tensor of the Majorana spinor.

 \end{abstract}

\vskip 1cm



\end{titlepage}

\newpage

\section{Introduction}

E.Majorana considered in a historic paper in 1937,a derivation of the Dirac equation where the electrons and the positrons are treated in a symmetric way\cite{majorana}.
He  made use of a self-conjugate representation of the $\gamma$-matrices that we now call the Majorana realization.
Furthermore he treated  the  fermion fields  in terms of anti-commuting variables. Such a theory has no
conventional classical interpretation. However, it makes it possible to derive the Dirac equation by a variational principle.
In a particular Majorana realization of $\gamma$-matrices as generators of the real part of the Cllfford algebra $Clif(1,3)$ over space-time,
all components of the 4-spinors turn out to be real. As such they correspond to electrically neutral fermion fields
whose quantization may be achieved on a real Hilbert space\cite{myrheim,arodz1,arodz2}.
\\
A candidate constituent particle for a Majorana spinor  is a neutrino\cite{senjanovic}.
It should be noted in the present approach that there is no need to distinguish a separate anti-neutrino of opposite chirality: a neutrino coming out together with an  electron in $\beta$-decay
would have negative helicity while that coming out together with a positron in inverse $\beta$-decay would have positive helicity.
In fact as already noted by Majorana himself, the Majorana nature of neutrino can be tested in {\it neutrino-less double $beta$ decays}\cite{CUORE}.
Such processes violate the lepton number
conservation by terms proportional to the Majorana mass of the neutrino.
Thus the Majorana nature of a neutrino emerges as the most natural explanation for the surprisingly small observed value of  neutrino masses. Furthermore
  the non-conservation of the lepton number $L$   leads to speculations that the decay of supermassive Majorana neutrinos in the very early Universe may have given rise to an
  asymmetry in $L$ that transforms to the present-day observed baryon number $B$ asymmetry by virtue of  the $B-L$ conservation\cite{fujikawa2}.
Several laboratories around the world support experiments to detect neutrino-less double $\beta$ decay, with no success up till now.
A new generation of experiments called CUORE is on its way in Italy to resolve this question.
A possible discovery would also provide a natural answer to the dominance of matter over anti-matter in our universe\cite{sterile neutrinos}.

\noindent  Our main goal in this paper is to provide a consistent derivation of the coupled field equations of a Majorana neutrino and Einstein's theory of gravitation.
We make essential use in our calculations of the anti-commuting nature of the spinor components.
Einstein's theory of general relativity is a theory of gravitation determined by the semi-Riemannian geometry of a 4-dimensional spacetime.
The coupled field equations of the theory may be obtained by field variational principle from an action. It is well known that the second order principle where the metric variations of the Levi-Civita connections are taken into account; and the first order (Palatini) where independent variations of the action relative to the metric and connection yield, in these cases, the same set of field equations.The fact that the connection is Levi-Civita may be imposed by the method of Lagrange multipliers.
In the following, both the first and the second order variational  approaches are given and the resulting
field equations are derived and then compared.

\newpage

\section{The Action}

\noindent The field equations will be derived from the action
\ba
I[e,\omega,\psi,\lambda] = \int_M {\mathcal{L}}
\ea
where the Lagrangian density 4-form is given by
\ba
{\mathcal{L}} =\frac{1}{2} R_{a b} \wedge *(e^a \wedge e^b) + \frac{i}{2} \bar{\psi} *\gamma \wedge \nabla \psi +\frac{i}{2} m (\bar{\psi} \psi)*1 .
\ea
$m$ is the Majorana mass. The basic field variables are the $g$-orthonormal basis 1-forms $\{e^a | a=0,1,2,3 \}$ in terms of which the spacetime metric reads
\ba
g = \eta_{ab} e^a \otimes e^b = -e^0 \otimes e^0 + e^1 \otimes e^1 + e^2 \otimes e^2 + e^3 \otimes e^3;
\ea
the metric-compatible connection 1-forms $\{ \omega^{a}_{\;\;b}\}$ so that  $ \omega_{ab} = -\omega_{ba}.$ The Cartan structure equations
\ba
T^a = de^a + \omega^{a}_{\;\;b} \wedge e^b
\ea
and
\ba
R^{a}_{\;\; b} = d\omega^{a}_{\;\; b} + \omega^{a}_{\;\; c} \wedge \omega ^{c}_{\:\: b}
\ea
determines the torsion 2-forms  $\{T^a\}$ and the curvature 2-forms $\{R^{a}_{\;\;b}\}$,respectively.
The Hodge $*$-map that takes $p$-forms to $(4-p)$-forms, $0 \leq p \leq 4$, is fixed for a particular choice of the space-time orientation by the invariant volume 4-form
  \ba
  *1 = e^0 \wedge e^1 \wedge e^2 \wedge e^3 = \frac{1}{4!} \epsilon_{abcd} \; e^a \wedge e^b \wedge e^c \wedge e^d .
\ea
We are going to employ a Majorana (real) realization of the $\gamma$-matrices $\{\gamma_a\}$ and let $\gamma = \gamma_a e^a.$ We set
\ba
\gamma_0 = \left ( \begin{array}{cc} 0&-\sigma_1 \\ \sigma_1 & 0 \end{array} \right ), \gamma_1 = \left ( \begin{array}{cc} I&0 \\ 0 &-I \end{array} \right ),
\gamma_2 = \left ( \begin{array}{cc} 0&-i\sigma_2 \\ -i\sigma_2 & 0 \end{array} \right ), \gamma_3 = \left ( \begin{array}{cc} 0&I \\ I& 0 \end{array} \right ),   \nonumber
\ea
so that
\ba
\gamma_5 = \gamma_0 \gamma_1 \gamma_2 \gamma_3 =  \left ( \begin{array}{cc} 0&\sigma_3 \\ -\sigma_3 & 0 \end{array} \right ).
\ea
The charge conjugation matrix $\mathcal{C}$ should satisfy
\ba
\mathcal{C}^{-1} = \mathcal{C}^{T}, \; \: \mathcal{C}^2 = -I, \: \:  \mathcal{C} \gamma_a \mathcal{C}^{-1} = {\gamma_a}^{T} .
\ea
We made the choice $ \mathcal{C} = \gamma_0$ in our Majorana realization. Given a 4-spinor $\psi$, its charge conjugated spinor is defined to be
\ba
\psi_{C} \equiv \mathcal{C} {\bar{\psi}}^{T}
\ea
 where $\bar{\psi} = \psi^{\dagger}\mathcal{C}.$ A Majorana 4-spinor is by definition a self-charge conjugate 4-spinor:
 \ba
 \psi_C = \psi.
 \ea
 In the Majorana realization of $\gamma$-matrices above, a Majorana spinor is a real spinor:
 \ba
 \psi_{Majorana} = \left ( \begin{array}{c} \psi_1 \\ \psi_2 \\ \psi_3 \\ \psi_4 \end{array}  \right ) \in {\mathbf{R}}^4 .
 \ea
 We also note that in a coordinate patch $\{x^{\mu}\}$, the spinor components $\{ \psi_{\alpha} | \alpha =1,2,3,4\}$ are odd-Grassmann valued, real functions of all coordinates. That is they are each nilpotent and anti-commute among themselves:
 \ba
 \psi_{\alpha} \psi_{\beta} + \psi_{\beta} \psi_{\alpha} = 0.
 \ea

  \noindent The covariant exterior derivative of a Majorana 4-spinor
  \ba
  \nabla \psi = d\psi + \frac{1}{2} \omega^{ab} \sigma_{ab} \psi
  \ea
  where $\sigma_{ab} = \frac{1}{4} \left [ \gamma_a , \gamma_{b} \right ]$  are the Lie algebra
  generators of the spin cover $Spin(1,3)$ of the local Lorentz group $SO(1,3)$.  In particular
  \ba
  *\left ( \gamma \wedge *\nabla \right )  = (\gamma \cdot \nabla)
  \ea
  is the Dirac operator on the real spin bundle over space-time.

\section{Variational Field Equations}
\subsection{Zero-torsion Constrained, Second-Order Variations}

\noindent We impose the zero-torsion condition by the method of Lagrange multipliers so that the  connection 1-forms are going to be the Levi-Civita connection determined
up to local Lorentz transformations by the metric. To that end
we introduce a set $\{\lambda_a\}$ of Lagrange multiplier 2-forms upon whose variations the zero-torsion constraint is imposed on the geometry of space-time. Then the connection becomes the Levi-Civita connection $\{{\hat{\omega}}^{a}_{\:\: b} \}$. Let the constraint Lagrangian density 4-form be given by
\ba
{\mathcal{L}}_C = T^a \wedge \lambda_a.
\ea
Then variation of the total action density
\ba
{\mathcal{L}}_T = {\mathcal{L}} + {\mathcal{L}}_C
\ea
 reads up to a closed form as follows:
\begin{eqnarray}
\dot{\mathcal{L}}_T &=& {\dot{e}}^a \wedge \frac{1}{2} R^{bc} \wedge *(e_a \wedge e_b \wedge e_c)
+ \frac{1}{2} {\dot{\omega}}^{ab}  \wedge *(e_a \wedge e_b \wedge e_c)  \wedge T^c \nonumber \\ &+& {\dot{e}}^a \wedge \frac{i}{2} m(\bar{\psi}\psi)*e_a + {\dot{e}}^a \wedge  D\lambda_a
+ \frac{1}{2} {\dot{\omega}}^{ab}  \wedge (e_b \wedge \lambda_a - e_a \wedge \lambda_b ) \nonumber \\
&+& \frac{i}{2} {\dot{e}}^a \wedge \bar{\psi} *(e^{b} \wedge e_a)  \wedge \gamma_b \nabla \psi
-\frac{i}{4}{\dot{\omega}}^{ab}  \wedge  \bar{\psi} *\gamma \wedge \sigma_{ab} \psi \nonumber \\
&+& \frac{i}{2} \bar{\dot{\psi}} *\gamma \wedge \nabla \psi + \frac{i}{2} \bar{\psi} *\gamma \wedge \nabla \dot{\psi} +T^a \wedge {\dot{\lambda}}_a + im \bar{\dot{\psi}} \psi *1 .
\end{eqnarray}
(1) The variations of ${\dot{\lambda}}_a$ imposes the zero-torsion constraint
\ba
T^a =0
\ea
that implies that the connection is the unique Levi-Civita connection 1-forms ${\hat{\omega}}^{a}_{\:\:b}$.
The remaining variational equations should be be solved under this constraint. \\ (2) Einstein field equations follow from the variations ${\dot{e}}^a$
 of co-frame 1-forms:
 \ba
 -\frac{1}{2} {\hat{R}}^{bc} \wedge *(e_a \wedge e_b \wedge e_c) = \frac{i}{2} \bar{\psi} *(e^{b} \wedge e_a)  \wedge \gamma_b {\hat{\nabla}} \psi  + \frac{i}{2} m (\bar{\psi}\psi)*e_a
 +{\hat{D}}\lambda_a.
 \ea

\noindent (3) The Dirac equation satisfied by the Majorana 4-spinor is determined from the variations of the Lagrangian density 4-form with respect to $\psi$. The relevant terms subject to the zero-torsion constraint are
\ba
\frac{i}{2} \bar{\dot{\psi}} *\gamma \wedge \hat{\nabla} \psi + \bar{\psi} *\gamma \wedge \hat{\nabla} \dot{\psi}.
\ea
We open up the covariant derivative in  the second term and use Majorana flip identities to write it as
\ba
\bar{\psi} *\gamma \wedge \hat{\nabla} \dot{\psi} = \frac{i}{2} \bar{d \dot{\psi}} \wedge *\gamma \psi -\frac{i}{2} {\hat{\omega}}^{ab} \wedge  \bar{\dot{\psi}} \sigma_{ab} *\gamma
\psi.
\ea
Then we differentiate the first term on the right by parts, use $\gamma$-matrix identities and the zero-torsion constraint to pass the covariant exterior derivative action
on $ \bar{\dot{\psi}}$ at the left to the covariant derivative action on $\psi$ at the right. Then up to a closed form we get the variation
\ba
i  \bar{\dot{\psi}} *\gamma \wedge \hat{\nabla} \psi .
\ea
Therefore the massless Dirac equation satisfied by the Majorana 4-spinor field follows:
\ba
i *\gamma \wedge \hat{\nabla} \psi + im \psi *1=0 .
\ea

\noindent  (4) The variation of the connections $\{\omega^{a}_{\;\;b}\}$ gives the field equations
\ba
e_a \wedge \lambda_b - e_b \wedge \lambda_a =\Sigma_{ab}
\ea
where
\ba
\Sigma_{ab} = -\frac{i}{2} \bar{\psi} *\gamma \sigma_{ab} \psi.
\ea
This is a system of linear algebraic equations that can be solved uniquely for the Lagrange multiplier 2-forms $\lambda_a.$ Let us
hit both sides with the interior product operators $\iota^a \equiv \eta^{ab} \iota_{X_b}$. Then
\ba
\lambda_a + e_a  \wedge \iota^b \lambda_{b} = \iota^b \Sigma_{ba}.
\ea
We hit this expression with $\iota^a$ once again so that
\ba
 \iota^a \lambda_{a} = \frac{1}{4} \iota^a \iota^b \Sigma_{ba}.
\ea
Now we use the $\gamma$-matrix identity
\ba
2\gamma_c \sigma_{ab} = \eta_{ca} \gamma_{b} - \eta_{cb} \gamma_a + \epsilon_{abcd} \gamma_{5} \gamma^{d};
\ea
the Majorana flip identity
\ba
\bar{\psi} \gamma_a \psi =0
\ea
satisfied by any odd-Grassmann valued (real) Majorana 4-spinor, and after a series of simplifications we arrive at the final expression
\ba
\lambda_a = \frac{i}{8} e_a \wedge (\bar{\psi} \gamma_5 \gamma \psi).
\ea
It is not difficult to verify that
\ba
\hat{D} \lambda_a =  \frac{i}{4} e_a \wedge (\bar{\psi} \gamma_5 \gamma \wedge \hat{\nabla}\psi).
\ea
Finally the Einstein field equations become
\ba
G_a = \frac{i}{2} \bar{\psi} *(e^{b} \wedge e_a)  \wedge \gamma_b {\hat{\nabla}} \psi  +  \frac{i}{4} e_a \wedge (\bar{\psi} \gamma_5 \gamma \wedge \hat{\nabla}\psi) +\frac{i}{2} m (\bar{\psi}\psi) *e_a ,
 \ea
where the Einstein 3-forms
\ba
G_a \equiv G_{ab} *e^b =  -\frac{1}{2} {\hat{R}}^{bc} \wedge *(e_a \wedge e_b \wedge e_c) .
\ea
Here the Einstein tensor $G= G_{ab} e^a \otimes e^b$ is symmetric by construction. We look at the two terms on the right hand side now.
The first term gives the canonical energy-momentum tensor $T[can] = T_{ab}[can] e^a \otimes e^b$ of the Majorana 4-spinor $\psi$:
\ba
\tau_{a}[can] \equiv T_{ab}[can] *e^b =   \frac{i}{2} \bar{\psi} *(e^{b} \wedge e_a)  \wedge \gamma_b {\hat{\nabla}} \psi +\frac{i}{2} m (\bar{\psi}\psi) *e_a .
\ea
The canonical energy-momentum tensor $T[can]$ is asymmetric in general. In fact we calculate its skew symmetric part as
\ba
e_a \wedge \tau_{b}[can] - e_b \wedge \tau_{b}[can] = \frac{i}{2} \bar{\psi} ( \eta_{ac} \gamma_b - \eta_{bc} \gamma_a )\psi *e^c .
\ea
On the other hand the skew symmetric part of the second term on the right hand side of the Einstein field equations reads
\ba
e_a \wedge \hat{D} \lambda_b   -e_b \wedge \hat{D} \lambda_a = \frac{i}{2} e_a \wedge e_b \wedge \bar{\psi} \gamma_5 \gamma \wedge \hat{\nabla}\psi.
\ea
Thus adding the last two equalities side by side and using the $\gamma$-matrix identity
\ba
2\sigma_{ab} \gamma_c = \eta_{ac}\gamma{b} - \eta_{bc} \gamma_a - \epsilon_{abcd} \gamma_5 \gamma^d,
 \ea
we prove that the skew symmetric part of the total energy-momentum tensor of a Majorana 4-spinor becomes
\ba
e_a \wedge \tau_{b}- e_b \wedge \tau_{b}= \frac{i}{2} \bar{\psi} \sigma_{ab} *\gamma \wedge \hat{\nabla}\psi .
\ea
Therefore the total energy-momentum tensor of $\psi$ is symmetric on-shell. Since $e_a \wedge \hat{D} \lambda_b  + e_b \wedge \hat{D} \lambda_a =0$ by inspection,
the symmetric part of the total energy-momentum tensor of $\psi$ is read from the symmetric part of the canonical energy-momentum tensor:
\begin{eqnarray}
e_a \wedge \tau_{b}[can] + e_b \wedge \tau_{b}[can] &=& i \eta_{ab} \bar{\psi} *\gamma \wedge \hat{\nabla}\psi + im \eta_{ab} (\bar{\psi}\psi) *1 \nonumber \\ & &
- \frac{i}{2} \bar{\psi} (*e_b \gamma_a + *e_a \gamma_b) \wedge \hat{\nabla}\psi.
\end{eqnarray}
Then noting that $*e_a \wedge \hat{\nabla}\psi = -\hat{\nabla}_{X_a} \psi *1$, and working on-shell we determine
\ba
T_{ab}[sym] = \frac{i}{2} \bar{\psi} (\gamma_a   \hat{\nabla}_{X_b}   + \gamma_b \hat{\nabla}_{X_a}  ) \psi + im \eta_{ab} (\bar{\psi} \psi)  .
\ea
as expected.

 \subsection{Non-constrained First-Order Variations}

 \noindent In this case the variations of the action density reads
\begin{eqnarray}
\dot{\mathcal{L}} &=& {\dot{e}}^a \wedge \frac{1}{2} R^{bc} \wedge *(e_a \wedge e_b \wedge e_c)
+ \frac{1}{2} {\dot{\omega}}^{ab}  \wedge *(e_a \wedge e_b \wedge e_c)  \wedge T^c \nonumber \\
&+& \frac{i}{2} {\dot{e}}^a \wedge \bar{\psi} *(e^{b} \wedge e_a)  \wedge \gamma_b \nabla \psi +  \frac{i}{2} {\dot{e}}^{a} \wedge m(\bar{\psi}\psi) *e_a
\nonumber \\ &-& \frac{i}{4}{\dot{\omega}}^{ab}  \wedge  \bar{\psi} *\gamma \wedge \sigma_{ab} \psi   \nonumber \\
&+& \frac{i}{2} \bar{\dot{\psi}} *\gamma \wedge \nabla \psi + \frac{i}{2} \bar{\psi} *\gamma \wedge \nabla \dot{\psi} + im \bar{\dot{\psi}} \psi *1.
\end{eqnarray}
The co-frame variations give the Einstein field equations
  \ba
 -\frac{1}{2} R^{bc} \wedge *(e_a \wedge e_b \wedge e_c) = \frac{i}{2} \bar{\psi} *(e^{b} \wedge e_a)  \wedge \gamma_b \nabla \psi + im (\bar{\psi} \psi) *e_a.
 \ea
 while the variations of the connection 1-forms provide the field equations satisfied by the torsion 2-forms:
 \ba
  *(e_a \wedge e_b \wedge e_c)  \wedge T^c  = \frac{i}{2} \bar{\psi} *\gamma \wedge \sigma_{ab} \psi .
  \ea
These are linear algebraic equations that determine the torsion 2-forms uniquely as
\ba
T^a = \frac{i}{4} *e^{a}_{\;\;b} (\bar{\psi} \gamma_5 \gamma^{b} \psi) .
\ea
 Then the contortion 1-forms will be
 \ba
 K^{a}_{\;\;b} = \frac{i}{8} *e^{a}_{\;\; bc} (\bar{\psi}\gamma_5 \gamma^{c} \psi).
 \ea
 The simplification of the Dirac equation needs some further manipulations.
 Variation of the action density gives
\ba
\frac{i}{2}\bar{\dot{\psi}} *\gamma \wedge \nabla \psi + \frac{i}{2} \bar{\psi} *\gamma \wedge ( d \dot{\psi} + \frac{1}{2} \omega^{ab} \sigma_{ab} \dot{\psi}).
\ea
We use Majorana flip identities on the second term and get
 \ba
\frac{i}{2}\bar{\dot{\psi}} *\gamma \wedge \nabla \psi + \frac{i}{2} d\bar{\dot{\psi}} \wedge *\gamma \psi - \frac{i}{4} \bar{\dot{\psi}} \sigma_{ab} *\gamma \psi \wedge \omega^{ab} .
\ea
 We now differentiate the second term by parts and use $\gamma$-matrix identities on the third term and get our expression simplified to ( up to a closed form)
\ba
i\bar{\dot{\psi}} *\gamma \wedge \nabla \psi + \frac{i}{2} \bar{\dot{\psi}} *e^{a}_{\;\; b} \gamma_a \psi  \wedge T^b + (closed  \;  form).
\ea
Therefore our Majorana 4-spinor field $\psi$ satisfies  the variational (massless) Dirac equation
\ba
i *\gamma \wedge \nabla \psi + \frac{i}{2} T^a \wedge *e_{a}^{\;\; b} \gamma_b \psi +  im \psi *1 = 0.
\ea

\medskip

\noindent  We now note that  connection 1-forms with torsion can be uniquely decomposed as the sum of the Levi-Civita connection 1-forms, determined only by the metric and the contortion 1-forms determined by our Majorana 4-spinor:
\ba
\omega^{a}_{\;\; b} = {\hat{\omega}}^{a}_{\;\;b} + K^{a}_{\;\;b}.
\ea
Then the curvature 2-forms can be similarly decomposed
\ba
R^{a}_{\;\; b} ={\hat{R}}^{a}_{\;\; b} + {\hat{D}}K^{a}_{\;\; b} + K^{a}_{\;\; c} \wedge K^{c}_{\;\; b}.
\ea
Therefore the Einstein field equations can be written as
\begin{eqnarray}
-\frac{1}{2} {\hat{R}}^{bc} \wedge *e_{abc} &=&  \frac{i}{2} \bar{\psi} *(e^{b} \wedge e_a)  \wedge \gamma_b \hat{\nabla} \psi + \frac{i}2 m (\bar{\psi}\psi)*e_a  + \frac{1}{2} {\hat{D}}K^{bc} \wedge *e_{abc}  \nonumber \\
&+&\frac{i}{4} \bar{\psi} *(e^{b} \wedge e_a)  \wedge \gamma_b K^{cd} \sigma_{cd} \psi + \frac{1}{2} K^{b}_{\;\; d} \wedge K^{dc} \wedge *e_{abc}.
\end{eqnarray}
It is not difficult to check that
\ba
  \frac{1}{2} {\hat{D}}K^{bc} \wedge *e_{abc} =  \frac{i}{4} e_a \wedge (\bar{\psi} \gamma_5 \gamma \wedge \hat{\nabla}\psi).
\ea
Therefore the first, second and third terms on the right hand side above add up to the symmetric stress-energy-momentum 3-forms of the Majorana spinor.
The sum of the fourth and fifth terms give a particular quartic self coupling of the Majorana spinor. We explicitly work it out to be
\ba
\frac{3}{64}(\bar{\psi} \gamma_5 \gamma_b \psi) (\bar{\psi} \gamma_5 \gamma^b \psi) *e_a .
\ea
In a similar way the Dirac equation decomposes as
\ba
i *\gamma \wedge \hat{\nabla} \psi +im \psi *1+ \frac{3}{16} (\bar{\psi} \gamma_5 \gamma \psi) \wedge \gamma_5 *\gamma \psi = 0.
\ea


 \section{Concluding Comments}

 \medskip

\noindent (A)  As far as the coupled field equations are concerned the zero-torsion constrained variations of the action density 4-form
 \begin{eqnarray}
 {\mathcal{L}} &=&\frac{1}{2} R_{a b} \wedge *(e^a \wedge e^b) + \frac{i}{2} \bar{\psi} *\gamma \wedge \nabla \psi +\frac{i}{2}m(\bar{\psi}\psi)*1 \nonumber \\
& &+ \frac{\alpha}{4} (\bar{\psi} \gamma_5 \gamma \psi) \wedge (\bar{\psi} \gamma_5 *\gamma \psi)  + T^a \wedge \lambda_a,
 \end{eqnarray}
 and the unconstrained (i.e. with non-vanishing torsion) variations of the action density 4-form
 \begin{eqnarray}
 {\mathcal{L}}^{\prime} &=& \frac{1}{2} R_{a b} \wedge *(e^a \wedge e^b) + \frac{i}{2} \bar{\psi} *\gamma \wedge \nabla \psi +\frac{i}{2}m(\bar{\psi}\psi)*1 \nonumber \\
& &+ \frac{\beta}{4} (\bar{\psi} \gamma_5 \gamma \psi) \wedge (\bar{\psi} \gamma_5 *\gamma \psi)
 \end{eqnarray}
yield  the same field equations provided their parameters are related by
 \ba
 \alpha=\beta  + \frac{3}{16}.
 \ea
 Of course the coupling of spinning matter to space-time geometry in these two cases may differ and needs further considerations.
 \medskip

\noindent  (B) A short remark concerning the physical interpretation of our Lagrange multiplier 2-forms is in order. Consider a generic matter Lagrangian density 4-form
\ba
{\mathcal{L}}_{matter}(\psi, e, \hat{\omega})
  \ea
 whose gravitational couplings are described through  its dependence on both the co-frame and the connection. The coupling of Dirac, Weyl or Majorana spinors to gravity provide familiar examples of such couplings.  On the other hand non-minimal couplings of electromagnetic fields to gravity described by generic terms of the type $RF^2$ in the Lagrangian density 4-form
 provide rather unfamiliar examples. Then the variational derivatives
 \ba
 \frac{\delta {\mathcal{L}}_{matter}}{\delta e^a} \equiv \tau_{a}(matter)
 \ea
 give the energy-momentum 3-forms of matter for which the corresponding canonical energy-momentum tensor is non-symmetric in general. Similarly
  \ba
 \frac{\delta {\mathcal{L}}_{matter}}{\delta {\omega}^{ab}} \equiv \Sigma_{ab}(matter)
 \ea
 give the angular momentum 3-forms of matter. We construct the 2-forms
 \ba
 \lambda_a  =\iota^b \Sigma_{ba} -\frac{1}{4} e_a \wedge \iota^c \iota^b \Sigma_{bc}
 \ea
 so that
 \ba
 \hat{D}\lambda_a = \hat{D}(\iota^b \Sigma_{ba}) + \frac{1}{4} e_a \wedge d(\iota^c \iota^b \Sigma_{bc}).
 \ea
 Then the energy-momentum tensor corresponding to the sum
 \ba
 \tau_{a}(matter) +   \hat{D}\lambda_a
 \ea
 is symmetric. This is the  {\sl Belinfante-Rosenfeld procedure} \cite{belinfante1,rosenfeld}for the construction of a symmetric energy-momentum tensor when the matter Lagrangian density 4-form carries an explicit dependence on the connection 1-forms. We will not repeat the proof that the conservation laws associated with both the canonical tensors and the symmetric tensors
 remain the same\cite{forger-romer} .
 Our main  observation here is  that our Lagrange multiplier 2-forms $\lambda_a$ are nothing but the well-known Belinfante-Rosenfeld 2-forms.

\medskip

 \noindent (C) Under local Pauli-G\"{u}rsey (chiral) rotations
 \ba
 \psi \rightarrow e^{\theta \gamma_5}  \psi,
 \ea
 there exists an axial vector current density
 \ba
 J_5 = \frac{i}{4} (\bar{\psi} \gamma_5 \gamma \psi)
 \ea
 such that
 \ba
 *d*J_5 =im (\bar{\psi} \gamma_5 \psi)
 \ea
 on-shell. $J_5$ is conserved for a massless-neutrino. The generic quartic self coupling of  Majorana neutrinos takes the form
 $  J_5 \wedge *J_5. $


\section{Acknowledgement}
T.D is partially supported by The Turkish Academy of Sciences (T\"{U}BA).


\end{document}